\pdfoutput=1 
\documentclass[aps,prl, amsmath,amssymb,superscriptaddress,showkeys,reprint,preprintnumbers,floatfix]{revtex4-2}
\usepackage{color}
\usepackage{aas_macros}
\usepackage{hyperref,breakurl}
\hypersetup{colorlinks=true, citecolor=blue, urlcolor=blue, linkcolor=blue}
\usepackage[usenames,dvipsnames]{xcolor}
\usepackage{dcolumn}
\usepackage{bm}
\usepackage{array}
\usepackage{comment}
\usepackage{makecell}
\usepackage{amsmath,amssymb,latexsym,times}
\usepackage{orcidlink}

\usepackage{todonotes}
\usepackage[normalem]{ulem}

\newcommand{\twolineconstraint}[3]{
\begin{equation}
\left.
\begin{aligned}
#1 \\ #2
\end{aligned}
\,\right\}\;\begin{tabular}{@{}l@{}}#3\end{tabular}
\end{equation}
}

\usepackage{natbib}
\usepackage{ifthen}

\defcitealias{y6-3x2pt}{\textsc{Y6-3$\times$2pt}}
\defcitealias{y6-extensions}{\textsc{Y6-Ext}}

\begin{document}

\title{Constraints on Dynamical Dark Energy from Multiple Probes in the Full Dark Energy Survey}


\author{T.~M.~C.~Abbott\,\orcidlink{0000-0003-1587-3931}}
\affiliation{Cerro Tololo Inter-American Observatory, NSF's National Optical-Infrared Astronomy Research Laboratory, Casilla 603, La Serena, Chile}

\author{M.~Adamow\,\orcidlink{0000-0002-6904-359X}}
\affiliation{Center for Astrophysical Surveys, National Center for Supercomputing Applications, 1205 West Clark St., Urbana, IL 61801, USA}

\author{M.~Aguena\,\orcidlink{0000-0001-5679-6747}}
\affiliation{INAF-Osservatorio Astronomico di Trieste, via G. B. Tiepolo 11, I-34143 Trieste, Italy}
\affiliation{Laborat\'orio Interinstitucional de e-Astronomia - LIneA, Av. Pastor Martin Luther King Jr, 126 Del Castilho, Nova Am\'erica Offices, Torre 3000/sala 817 CEP: 20765-000, Brazil}

\author{A.~Alarcon\,\orcidlink{0000-0001-8505-1269}}
\affiliation{Institute of Space Sciences (ICE, CSIC),  Campus UAB, Carrer de Can Magrans, s/n,  08193 Barcelona, Spain}

\author{S.~Allam\,\orcidlink{0000-0002-7069-7857}}
\affiliation{Fermi National Accelerator Laboratory, P. O. Box 500, Batavia, IL 60510, USA}

\author{O.~Alves}
\affiliation{Kavli Institute for the Physics and Mathematics of the Universe (WPI), UTIAS, The University of Tokyo, Kashiwa, Chiba 277-8583, Japan}
\affiliation{Department of Physics, University of Michigan, Ann Arbor, MI 48109, USA}

\author{A.~Amon\,\orcidlink{0000-0002-6445-0559}}
\affiliation{Department of Astrophysical Sciences, Princeton University, Peyton Hall, Princeton, NJ 08544, USA}

\author{D.~Anbajagane\,\orcidlink{0000-0003-3312-909X}}
\affiliation{Kavli Institute for Cosmological Physics, University of Chicago, Chicago, IL 60637, USA}

\author{F.~Andrade-Oliveira\,\orcidlink{0000-0003-0171-6900}}
\affiliation{Physik-Institut, University of Zürich, Winterthurerstrasse 190, CH-8057 Zürich, Switzerland}

\author{P.~Armstrong\,\orcidlink{0000-0003-1997-3649}}
\affiliation{The Research School of Astronomy and Astrophysics, Australian National University, ACT 2601, Australia}

% ...

\collaboration{DES Collaboration}

\noaffiliation

\date{\today}

\label{firstpage}

\begin{abstract}
We present results on dark energy evolution, assuming a time-dependent equation of state $w(a)=w_0+w_a(1-a)$, from growth and geometric probes using the full six-year Dark Energy Survey (DES) dataset: ~type Ia supernovae (SNe), baryon acoustic oscillations (BAO), and weak gravitational lensing and galaxy clustering (3$\times$2pt). The combination yields $w_0=-0.84^{+0.10}_{-0.10}$ and $w_a=-0.44^{+0.60}_{-0.55}$, the tightest constraints ever obtained from a single survey, and shows a $2.2\sigma$ deviation from a cosmological constant. Adding the Dark Energy Spectroscopic Instrument (DESI) DR2 BAO data yields $w_0=-0.84^{+0.06}_{-0.07}$ and  $w_a=-0.53^{+0.33}_{-0.28}$, representing the most stringent low-redshift-only test of dynamical dark energy to date, with a  $2.3\sigma$ deviation. In this combination, the inclusion of 3$\times$2pt doubles the constraining power. Finally, when combined with primary CMB information from a combination of \textit{Planck}, the Atacama Cosmology Telescope, and the South Pole Telescope, we obtain  $w_0=-0.82^{+0.05}_{-0.05}$, $w_a=-0.63^{+0.21}_{-0.18}$,  $3.0\sigma$ away from a cosmological constant. We find that including 3$\times$2pt in the previously studied SN + DESI BAO + CMB combination leaves the significance essentially unchanged ($3.2 \sigma$ to $3.0\sigma$) while improving the figure of merit by $\sim$10\%. We systematically investigate the impact of leaving out each one of the probes (SN, BAO, 3$\times$2pt, CMB) and find that the significance of the deviation from a cosmological constant ranges from 2.3 to 3.2$\sigma$, with best-fit parameters always occupying the region where $w_0 >-1$ and $w_a <0$. Excluding SN from the all data combination yields a $2.6\sigma$ departure from $\Lambda$CDM, providing a cross-check independent of supernova photometric calibration. We find the residual $S_8$ parameter difference between the CMB and 3$\times$2pt datasets is unaffected by the assumption of dynamical dark energy. These results support the weak preference for evolving dark energy reported by several recent cosmological analyses. By combining growth and geometric probes from a single survey, this work realizes the multi-probe dark energy program envisioned at the inception of DES. 
\end{abstract}


\preprint{DES-2026-0979}
\preprint{FERMILAB-PUB-26-0306-PPD}

\maketitle

\section{Introduction}\label{sec:intro}

Understanding the mechanism driving the accelerated expansion of the Universe remains one of the most pressing open questions in modern cosmology. The cosmological constant, $\Lambda$, provides a simple and remarkably successful phenomenological description of dark energy, consistently fitting a wide range of cosmological observations over the past few decades~\cite{y1-keypaper, y3-3x2ptkp, y6-3x2pt, DESY5SN2024, Popovic2025a, Brout2022, Hoyt2026, Planck2020_cosmo, ACTDR6, SPT3G-DR1}.
However, its physical interpretation remains deeply problematic: if $\Lambda$ is interpreted as the vacuum energy density of spacetime, its observed value is many orders of magnitude smaller than naive expectations from quantum field theory, giving rise to the well-known cosmological constant problem~\cite{Weinberg1989, Jerome2012}; if some physical principle sets the fundamental vacuum energy density to zero or to a value much smaller than the present energy density of the Universe, then it is plausible that dark energy arises as a dynamical phenomenon associated with novel fields not yet having reached their ground states \cite{Frieman1995}.

A key question is therefore whether the dark energy density is truly a constant or instead evolves with time.  
Detecting any time dependence would provide crucial insight into 
the underlying nature of dark energy 
potentially pointing to new dynamical degrees of freedom. A wide class of theoretical models predict such evolution, including scalar field scenarios such as quintessence~\cite{Frieman1995,Ratra1988, Caldwell1998}, k-essence~\cite{Chiba2000}, and phantom fields~\cite{Caldwell1999}, each characterized by a time-dependent equation of state $w(a)$. The simplest parameterization is a constant $w$ that is currently found to be consistent with $-1$~\cite{y6-3x2pt}. Going beyond that, a commonly used framework to capture such dynamics at low redshifts is the Chevallier–Polarski–Linder (CPL) parameterization~\cite{chevallier_polarski, linder_w0wa} (referred to as $w_0 w_a$CDM): 
\begin{equation}
w(a) = w_0 + w_a(1-a)\,.
\end{equation}
Recent observations of Type Ia supernovae (SNe) \cite{DESY5SN2024, Popovic2025b} and Baryon Acoustic Oscillations (BAO) \citep{desi-dr1, y6-baokp}  have provided intriguing hints of dynamical dark energy, which also persist when combined with Cosmic Microwave Background (CMB) data, with a significance at the $2$--$4\sigma$ level within the CPL framework~\cite{DESY5SN2024,desi-dr2,DESI:2025fii,des-bao-sn-cosmo,Popovic2025a}.
Notably, the best-fit results in the context of the CPL parametrization suggest a transition in the equation of state parameter, with $w<-1$ at high redshift and $w>-1$ at low redshift. Such a ``phantom-crossing" behavior~\cite{Hu:2004kh,Guo:2004fq} cannot be realized within canonical single-field models and instead points toward more complex scenarios, such as non-minimal couplings between dark energy and dark matter~\cite{Khoury:2025txd,Bedroya:2025fwh}, or modifications of gravity~\cite{Ye:2024ywg, Pan:2025psn,Wolf:2025jed,Wolf:2025acj,Tsujikawa:2025wca, Cataneo:2025vae}. In turn, these possibilities offer an avenue for probing the microphysics of dark energy. 

The Dark Energy Survey (DES) \cite{DES2016} 
full six-year dataset (DES Y6) \cite{des-dr2} enables an independent cross-check of such claims with unprecedented precision. In addition to BAO and SN data, which were used to explore constraints on dynamical dark energy in Ref.~\cite{des-bao-sn-cosmo}, here we also include the DES Y6 3$\times$2pt measurements, which are based on a series of papers \cite{y6-gold, y6-balrog,y6-piff,y6-maglim,y6-imagesims, y6-metadetect,y6-mask,y6-sourcepz, y6-lenspz,y6-wz,y6-nzmodes,y6-magnification,y6-cardinal, y6-gglens,y6-methods,y6-ppd,y6-1x2pt,y6-3x2pt}. When combined with BAO and SN, 3$\times$2pt breaks degeneracies in parameter space and provides an independent assessment of dynamical dark energy without including CMB information. This is  the most precise measurement to date of dark energy from a single survey. In addition, we also combine our DES-only results with external geometric and CMB constraints.

The full analysis details associated with this Letter, as well as additional extensions to the $\Lambda$CDM model, will be presented in an upcoming paper \citep[][hereafter \citetalias{y6-extensions}]{y6-extensions}.

\section{Datasets}

\subsection{The Dark Energy Survey}
\label{sec:des}

\subsubsection{Weak Gravitational Lensing and Galaxy Clustering: 3$\times$2pt}
\label{sec:3x2pt}

The DES Y6 3$\times$2pt analysis jointly measures cosmic shear, galaxy-galaxy lensing, and galaxy clustering~\cite{y6-1x2pt,y6-gglens,y6-2x2pt} from two galaxy samples covering $\sim$4,300$~{\rm deg^2}$~\cite{y6-gold}: a shear sample of $\sim$140 million galaxies in four tomographic bins~\cite{y6-metadetect,y6-piff,y6-imagesims,y6-cardinal} up to $z \lesssim 2$ and a position sample of $\sim$9 million \textsc{MagLim++} galaxies in six bins over $0.2 < z < 1.2$~\cite{y6-maglim,y6-mask}.

The analysis described in this Letter largely follows the methodology in Ref.~\citep[][hereafter \citetalias{y6-3x2pt}]{y6-3x2pt}, 
and we adopt the same cosmological parameter priors as in Ref. \citepalias{y6-3x2pt}, except for the additional flat priors $w_0 \in [-3, -0.33]$ and $w_a \in [-3, 3]$,  with $w_0 + w_a < 0$. The main differences are: (1) we define our analysis choices and scale cuts taking into account the additional constraining power from external data, (2) consequently we need to exclude additional small scales from cosmic shear, (3) we adopt the simpler nonlinear alignment (NLA) model for intrinsic alignments instead of the Tidal Alignment and Tidal Torque (TATT), and (4) we fix the sum of neutrino masses to $\sum m_\nu = 0.06\,\mathrm{eV}$, to reduce projection effects in the marginalized posteriors.
The motivation for these changes is further discussed in Ref.~\citepalias{y6-extensions}, while in the next section we show that none of our results is driven by them. The blinding procedure follows that of  Ref.~\citepalias{y6-3x2pt}:  cosmological results were hidden until the analysis choices and robustness tests were finalized for this model, at which point the chains were unblinded without further modifications.

\subsubsection{Type Ia Supernovae}\label{sec:sne}

The DES supernova dataset comprises 1,623 photometrically identified type Ia supernovae (SNe) from the full DES five-year supernova survey \cite{Sanchez2024, DESY5SN2024}, combined with approximately 200 SNe from historical low-redshift samples \cite{Hicken2009, Hicken2012, Krisciunas2017, Foley2018}. In this work, we use DES-Dovekie~\cite{Popovic2025a}, a reanalysis of the original DES SNe data with updated photometric cross-calibration. While the results obtained before and after recalibration are consistent with each other, when combined with the primary CMB \citep{Planck2020_cosmo, ACTDR6, SPT3G-DR1} and the BAO measurements from DESI Data Release 2 \citep[DR2,][]{desi-dr2}, the recalibrated sample yields a reduced significance for dynamical dark energy ($3.2\sigma$) compared to the original DES-SN5YR dataset ($4.2\sigma$)~\cite{Popovic2025a}. 
Other SN catalogues, namely Pantheon+ and Union3.1, also lead to similar dynamical dark energy significances (of $3.2\sigma$ and $3.4\sigma$, respectively) after recent reanalyses, when combined with DESI DR2 BAO and different CMB combinations \cite{Hoyt2026}. 
At the time of this analysis, a parallel effort is in progress combining the DES-Dovekie and the Pantheon+ datasets \cite{SNUnite}.
We leave the combination of that new SN dataset and other probes to future analyses.

\subsubsection{Baryon Acoustic Oscillations}\label{sec:bao}

DES measures the baryon acoustic oscillation feature from the clustering of $\sim$16 million red and bright galaxies over 4,273~deg$^2$ in six tomographic bins in the redshift range $0.6 < z < 1.2$, at an effective redshift of $z_\mathrm{eff} = 0.851$~\cite{y6-baokp}. The combined measurement of angular-diameter distance to sound horizon scale of $D_M(z_\mathrm{eff})/r_d = 19.51 \pm 0.41$ represents a 2.1\% constraint, the tightest BAO measurement from a photometric survey to date. Since the DES footprint overlaps with DESI by $\sim$1,000~deg$^2$, a separate analysis excluding this overlap region yields $D_M(z_\mathrm{eff})/r_d = 19.74 \pm 0.60$~\cite{Mena26}, which we use when combining with DESI BAO to avoid double-counting information. When used without DESI BAO, the full DES BAO dataset is employed. Hereafter, we refer to the combination of DESI BAO and DES BAO (excluding the overlap region) simply as BAO.

\subsection{External Data}
\label{sec:ext_data}

To further constrain a dynamic dark energy, we also combine DES data with state-of-the-art measurements from other experiments.

We use DESI DR2 BAO \cite{desi-dr2}. The DESI data consist of ratios of distances, either parallel or transverse to the line-of-sight, to the sound horizon at the baryon drag epoch, $r_d$. These distances are measured from the two-point correlation function of several tracers of the matter distribution, including galaxies, quasars and the Lyman-$\alpha$ forest. The tracers' effective redshifts cover the range $0.295 < z < 2.330$.

Additionally, we consider CMB temperature and polarization (TT, EE and TE) anisotropy power spectra from the \textit{Planck} collaboration Public Release 3 (PR3)~\cite{Planck2020_cosmo}, the Atacama Cosmology Telescope DR6~\cite{ACTDR6}, and South Pole Telescope SPT-3G DR1~\cite{SPT3G-DR1}. These datasets are combined as described in Ref.~\citepalias{y6-3x2pt}, and the combination has been previously considered in Ref.~\cite{SPT3G-DR1}. 
We do not include CMB lensing since a proper combination would require modeling the cross-covariance with the 3$\times$2pt probe.

\section{Results}\label{sec:results}

\begin{figure}
\begin{center}
\includegraphics{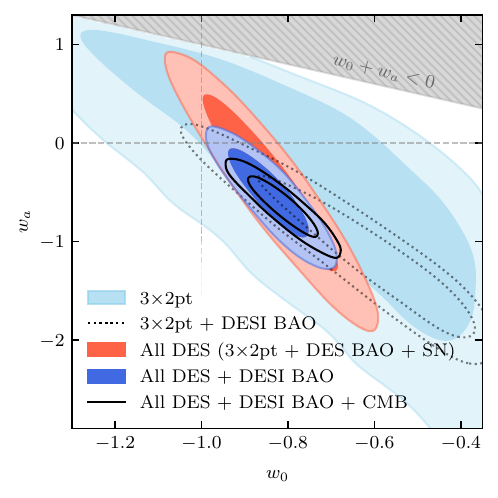}
\end{center}
\caption{
Constraints on the $w_0w_a$CDM parameters from DES 3$\times$2pt data (light blue), from DES 3$\times$2pt data and DESI BAO (dotted), from all DES probes (3$\times$2pt + SN + DES BAO; red filled), from all DES probes combined with DESI BAO (3$\times$2pt + SN + DES BAO + DESI BAO; blue filled), from all datasets considered in this work combined (3$\times$2pt + SN + BAO + CMB; black solid). The dashed crosshair marks the cosmological constant values ($w_0 = -1$, $w_a = 0$). The hatched region is excluded by the prior $w_0 + w_a < 0$.
All contours show 68\% and 95\% credible regions.
}
\label{fig:w0wa}
\end{figure}

In this section, we present constraints on the $w_0$--$w_a$ parameter space. We use the Nautilus nested sampler \cite{Lange2023} for all parameter inference reported here. 
Unless otherwise stated, all parameter constraints are quoted as the posterior mean in each parameter with the 68\% credible level (C.L.) of posterior volume. 
In Fig.~\ref{fig:w0wa}, we show the posteriors in the $w_0$--$w_a$ parameter plane for multiple dataset combinations, including 3$\times$2pt. 
In Table~\ref{table:significance}, we report the distance to the cosmological constant case in units of $\sigma$, which we define based on the fraction of posterior mass enclosed by the contour that intersects the $(w_0, w_a) = (-1,0)$ point, following the posterior-difference method in Ref.~\citepalias{y6-3x2pt}. We also show, for each combination, its corresponding dark energy figure of merit, $\text{FoM} \equiv 1/\sqrt{\text{det}(\text{Cov}_{w_0w_a})}$, which measures the overall strength of the constraints in the $w_0$--$w_a$ plane \cite{DEFT,Wang2008}. We also report the $\Delta\chi^2$-based significance, with $\Delta \chi^2 \equiv \chi^2_{w_0w_a {\rm CDM}} - \chi^2_{\rm \Lambda CDM}$ computed at the maximum-a-posteriori (MAP) point of each model, which closely matches the one reported by 
the posterior-difference method. 
We have also checked that the evidence ratios corresponding to these data combinations show a picture consistent with the metrics listed in Table~\ref{table:significance}. 

The DES 3$\times$2pt data are consistent with a cosmological constant at the $1.0\sigma$ level. Marginalized constraints (95\% C.L.) yield $w_0 > -1.26$ and $w_a = -0.43^{+1.10}_{-0.78}$, representing an improvement of 22\% over DES Year~3 on $w_a$ \cite{y3-3x2pt-ext-kp}. 
For the first time, we combine 3$\times$2pt with DES BAO and SN measurements to provide a measurement of $w_0$ and $w_a$ parameters that 
uses only cosmological probes from a single survey.
We obtain:
\twolineconstraint
{w_0 &= -0.84 \pm 0.10}
{w_a &= -0.44^{+0.60}_{-0.55}}
{All DES (68\% C.L.)\\(3$\times$2pt + DES BAO + SN),}

\noindent a result that is $2.2\sigma$ away from the cosmological constant value $(w_0,w_a) = (-1, 0)$.
In this combination, 3$\times$2pt data improves the FoM by a factor of four compared to the purely geometric DES BAO + SN, showing the complementarity between those datasets.

When further combining our data with BAO measurements from DESI, we obtain: 
\twolineconstraint
{w_0 &= -0.84^{+0.06}_{-0.07}} 
{w_a &= -0.53^{+0.33}_{-0.28}}
{All DES + DESI BAO\\(68\% C.L.),}

\noindent which is $2.3\sigma$ away from $\Lambda$CDM. This result is still entirely based on low-redshift data, independent of the CMB. As can be seen in Table \ref{table:significance}, including 3$\times$2pt in this combination increases the FoM from 61 to 110, highlighting the importance of the 3$\times$2pt information for low-redshift tests of dark energy.
Adding CMB data to the low-redshift combination roughly doubles the FoM (from 110 to 222), owing to the long distance lever arm to recombination.
With the combination 
All DES + 
DESI BAO + CMB  we obtain:
\twolineconstraint
{w_0 &= -0.82 \pm 0.05}
{w_a &= -0.63^{+0.21}_{-0.18}}
{All DES + DESI BAO + CMB\\(68\% C.L.),}

\noindent which is $3.0\sigma$ away from $\Lambda$CDM.
Compared to the SN + BAO + CMB combination, the addition of 3$\times$2pt improves the FoM from 202 to 222 ($10\%$), 
while the change in the $\Lambda$CDM-departure significance ($3.2\sigma$ to $3.0\sigma$) is 
consistent with the agreement we find between our two significance metrics (see Table \ref{table:significance}).

 \begingroup
 \renewcommand{\arraystretch}{1.2} 
 \begin{table}
 \caption{Distance to $\Lambda$CDM in the $w_0w_a$CDM parameter space for different dataset combinations, defined as the fraction of posterior mass enclosed by the isolikelihood contour that intersects $(w_0, w_a) = (-1,0)$. 
 We also show $\Delta\chi^2$-based model comparison, defined as $\Delta \chi^2 \equiv \chi^2_{w_0w_a {\rm CDM}} - \chi^2_{\rm \Lambda CDM}$ computed at the maximum-a-posteriori (MAP) point of each model. 
 Significances from both metrics are reported in units of Gaussian standard deviations, following Ref.~[\citetalias{y6-3x2pt}].
The last column shows the dark energy figure of merit $\text{FoM} \equiv 1/\sqrt{\text{det}(\text{Cov}_{w_0w_a})}$. We refer to the combination of DESI BAO and DES BAO (excluding the area overlap, see text for details) as BAO, while SN corresponds to the DES Dovekie recalibrated sample. }
 \label{table:significance}
 \begin{center}
 \begin{tabular}{lccc }
 \hline
 \hline
 Datasets & \begin{tabular}[c]{@{}c@{}}Distance to\\   $\Lambda$CDM ($\sigma$)\end{tabular} & \begin{tabular}[c]{@{}c@{}}$\Delta\chi^2$ \\ ($\sigma$)\end{tabular} &  FoM \\ \hline
 3$\times$2pt                                                     & $1.0$ & {$1.0$} & $6$ \\
 SN + DES BAO                                                     & $1.5$ &  {$1.6$} & $11$\\
 \textbf{All DES } (3$\times$2pt + SN + DES BAO)                & $2.2$ &  {$2.0$} & $48$  \\
 \textcolor{black}{All data} (All DES + DESI BAO + CMB)           & $3.0$ &  {$3.0$}& $222$ \\ \hline
 SN + BAO                                                         & $2.4$ &  {$2.3$}& $61$  \\
 3$\times$2pt + DESI BAO                                          & $1.7$ &  {$1.5$} & $32$  \\
 CMB + DESI BAO                                                   & $2.9$ &  {$2.9$} & $61$  \\ \hline
 All DES + DESI BAO ({All except CMB})      & $2.3$ &  {$2.2$} & $110$ \\
 SN + BAO + CMB ({All except 3$\times$2pt})        & $3.2$ &  {$3.3$} & $202$ \\
 3$\times$2pt + BAO + CMB ({All except SN}) & $2.6$ & {2.5} & $66$  \\
 3$\times$2pt + SN + CMB ({All except BAO}) & $2.4$ & {2.4} & $96$  \\
 All DES + CMB ({All except DESI BAO})      & $2.8$ & {$2.8$}  & $102$  \\ \hline
 \hline
 \end{tabular}
 \end{center}
 \end{table}
 \endgroup

In the lower part of Table~\ref{table:significance}, we investigate the robustness of our results by systematically removing one probe type at a time from the full dataset combination, and report the distance to $\Lambda$CDM and the FoM for each new combination, similar to the study in Ref.~\cite{Ishak2025}.
In each case, the remaining data show varying levels of difference from $\Lambda$CDM, 
ranging from $2.3\sigma$ (excluding CMB) to $3.2\sigma$ (excluding 3$\times$2pt). 
These variations are stable in terms of their central values, 
with best-fit parameters always occupying 
the region of $w_0 > -1$ and $w_a < 0$. 
Thanks to the improved constraining power of the DES Y6 3$\times$2pt, we are also able to perform an inference leaving out \textit{both} SN and CMB. The combination of 3$\times$2pt + DESI BAO deviates from $\Lambda$CDM at $1.7\sigma$, providing a test of dynamical dark energy independent of SN and CMB.

\begin{figure}
    \centering
    \includegraphics[width=0.49\textwidth]{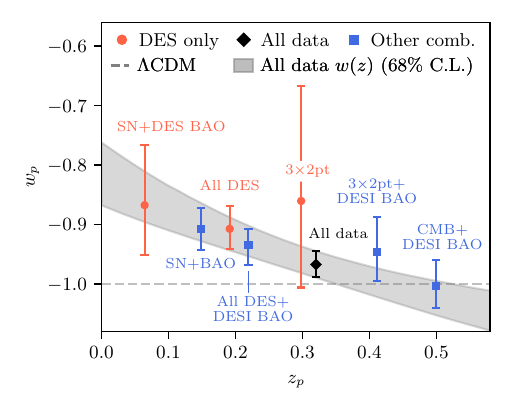}
    \caption{The best-constrained values of the pivot $w_p$ are shown at their pivot redshift for a variety of combinations of probes studied in this paper. The 68\% C.L. of $w(z)$ obtained using the CPL parametrization is shown as the shaded region for the combination of all data. 
    The DES data, now including 3$\times$2pt, allow us to have multiple redundant combinations of probes that can constrain $w_p$ at a range of pivot redshifts spanning much of the late-time expansion history.} 
    \label{fig:wp}
\end{figure}

In Fig.~\ref{fig:wp}, we summarize the best-constrained values of the pivot $w_p$ 
from the various dataset combinations explored in this work. 
The pivot value is defined as $w_p \equiv w(z_p)$, where the pivot redshift $z_p$ is chosen such that $\mathrm{Cov}(w_p, w_a) = 0$ \cite{Huterer2001}. 
The DES data, including 3$\times$2pt, allow us to have multiple redundant combinations of probes that can constrain $w_p$ at a range of pivot redshifts spanning much of the late-time expansion history. 
Without the DES data, only the single CMB + DESI BAO point (rightmost) would be available.

We have also verified that allowing $\sum m_\nu$ to vary (assuming a uniform prior of $\sum m_\nu > 0.06 \, \mathrm{eV}$) does not significantly shift the $w_0$--$w_a$ constraints; the best-fit values and significance levels remain consistent with the fixed-mass case. Our tightest combination yields an upper bound of $\sum m_\nu < 0.18$~eV (95\% CL) with best-fit values $w_0=-0.80^{+0.06}_{-0.06}$ and $w_a=-0.74^{+0.25}_{-0.21}$, yielding a distance to a cosmological constant of $3.4 \sigma$. 
The slight increase from $3.0\sigma$ to $3.4\sigma$ reflects a shift in the central value of $w_a$ toward more negative values driven by the $\sum m_{\nu}$--$w_a$ degeneracy.

Finally, we test the robustness of the low-redshift-only case (All DES + DESI BAO) and 
the combination of all data 
to alternative analysis choices, including different intrinsic alignment models (TATT and NLA with free amplitude per redshift bin), freeing the baryonic feedback amplitude, and including the second tomographic bin of the position sample excluded from the baseline (see  Ref.~\citepalias{y6-extensions} for details). For each variant, we compute both the 1D shift in $w_0$, $w_a$, $\Omega_{\rm m}$, and $S_8$ relative to the baseline, and also the change in the significance of the difference of the fit from $\Lambda$CDM. We find that all parameter shifts are within $0.1\sigma$ in $w_0$ and $w_a$, within $0.2\sigma$ for $\Omega_{\rm m}$, and $S_8$, and that the distance to $\Lambda$CDM changes by no more than $\pm0.3\sigma$. 
These tests confirm that our results are not sensitive to these analysis choices.
We note that the second tomographic bin of the position sample, excluded from our baseline following Ref.~\citepalias{y6-3x2pt}, continues to show redshift uncertainty values lying outside its calibration prior when included, confirming that this issue persists in the $w_0w_a$CDM model.

Beyond the dark energy constraints, we also examine the implication for $S_8$ in the context of evolving dark energy. We find the $S_8$ difference between 3$\times$2pt and the primary CMB unchanged. The full analysis is presented in the Supplemental Material.

\section{Conclusion}\label{sec:conclusion}

In this Letter, we present constraints on dynamical dark energy from a combination of growth and geometric probes from the full six-year dataset from DES. 
These are the first constraints on the evolution of dark energy using multiple probes from a \textit{single} Stage III survey --- a major milestone for the experimental dark energy program.  This result combines one growth probe (3$\times$2pt) and two geometry probes (SN and BAO) from DES, achieving a dark energy FoM of 48. The marginalized values yield $w_0=-0.84^{+0.10}_{-0.10}$, $w_a=-0.44^{+0.60}_{-0.55}$, and is $2.2\sigma$ away from a cosmological constant.

Going beyond a single survey and combining the DES-only results with DESI DR2 BAO data yields the most stringent low-redshift-only test of dynamical dark energy to date. We find $w_0=-0.84^{+0.06}_{-0.07}$ and  $w_a=-0.53^{+0.33}_{-0.28}$. The inclusion of 3$\times$2pt in this combination doubles the constraining power. The constraining power doubles again when combining these datasets with primary CMB data. The final constraints are $w_0=-0.82^{+0.05}_{-0.05}$, $w_a=-0.63^{+0.21}_{-0.18}$,  $3.0\sigma$ away from a cosmological constant.

The redundancy of growth and geometry probes in our final combination also allows us to examine the sensitivity of our results when excluding one individual measurement at a time from the full combination. In these tests the posterior contours remain stable 
in location with best-fit parameters consistently in the region $w_0 > -1$, 
$w_a < 0$, and the significance varies from $2.3\sigma$ to $3.2\sigma$. The pivot values of $w$ from these redundant combinations of probes span a broad redshift range ($z_p \sim 0.05\text{–}0.5$), underscoring that these constraints probe the dark energy equation of state across much of the late-time expansion history.

This Letter marks a significant step towards completing the multi-probe dark energy program laid out in the early 2000s \cite{DEFT}, setting the stage for the next chapter of the program with Stage-IV imaging surveys such as Rubin, Euclid, and Roman.  

\label{lastpage}

\section*{Acknowledgements}~\label{sec:acknowledgements}

Funding for the DES Projects has been provided by the U.S. Department of Energy, the U.S. National Science Foundation, the Ministry of Science and Education of Spain, the Science and Technology Facilities Council of the United Kingdom, the Higher Education Funding Council for England, the National Center for Supercomputing Applications at the University of Illinois at Urbana-Champaign, the Kavli Institute of Cosmological Physics at the University of Chicago, the Center for Cosmology and Astro-Particle Physics at the Ohio State University, the Mitchell Institute for Fundamental Physics and Astronomy at Texas A\&M University, Financiadora de Estudos e Projetos, Funda{\c c}{\~a}o Carlos Chagas Filho de Amparo {\`a} Pesquisa do Estado do Rio de Janeiro, Conselho Nacional de Desenvolvimento Cient{\'i}fico e Tecnol{\'o}gico and the Minist{\'e}rio da Ci{\^e}ncia, Tecnologia e Inova{\c c}{\~a}o, the Deutsche Forschungsgemeinschaft and the Collaborating Institutions in the Dark Energy Survey. 

The Collaborating Institutions are Argonne National Laboratory, the University of California at Santa Cruz, the University of Cambridge, Centro de Investigaciones Energ{\'e}ticas, Medioambientales y Tecnol{\'o}gicas-Madrid, the University of Chicago, University College London, the DES-Brazil Consortium, the University of Edinburgh, the Eidgen{\"o}ssische Technische Hochschule (ETH) Z{\"u}rich, Fermi National Accelerator Laboratory, the University of Illinois at Urbana-Champaign, the Institut de Ci{\`e}ncies de l'Espai (IEEC/CSIC), the Institut de F{\'i}sica d'Altes Energies, Lawrence Berkeley National Laboratory, the Ludwig-Maximilians Universit{\"a}t M{\"u}nchen and the associated Excellence Cluster Universe, the University of Michigan, NSF NOIRLab, the University of Nottingham, The Ohio State University, the University of Pennsylvania, the University of Portsmouth, SLAC National Accelerator Laboratory, Stanford University, the University of Sussex, Texas A\&M University, and the OzDES Membership Consortium.

Based in part on observations at NSF Cerro Tololo Inter-American Observatory at NSF NOIRLab (NOIRLab Prop. ID 2012B-0001; PI: J. Frieman), which is managed by the Association of Universities for Research in Astronomy (AURA) under a cooperative agreement with the National Science Foundation.

The DES data management system is supported by the National Science Foundation under Grant Numbers AST-1138766 and AST-1536171. This work used Jetstream2 and OSN at Indiana University through allocation PHY240006: Dark Energy Survey from the Advanced
Cyberinfrastructure Coordination Ecosystem: Services $\&$ Support (ACCESS) program, which is supported by U.S. National Science Foundation grants
2138259, 2138286, 2138307, 2137603, and 2138296.

The DES Spanish institutions are partially supported by MICINN/MICIU/AEI (/10.13039/501100011033) under grants PID2021-123012NB, PID2021-128989NB, PID2022-141079NB, PID2023-153229NA, PID2024-159420NB, PID2024-156844NB, CEX2020-001058-M, CEX2020-001007-S and CEX2024-001441-S, some of which include ERDF/FEDER funds from the European Union, and a grant by LaCaixa Foundation (ID 100010434) code LCF/BQ/PI23/11970028. IFAE is partially funded by the CERCA program of the Generalitat de Catalunya.
 

Research leading to these results has received funding from the European Research
Council under the European Union's Seventh Framework Program (FP7/2007-2013) including ERC grant agreements 240672, 291329, and 306478.
We  acknowledge support from the Australian Research Council Centre of Excellence for All-sky Astrophysics (CAASTRO), through project number CE110001020, and the Brazilian Instituto Nacional de Ci\^encia
e Tecnologia (INCT) e-Universe (CNPq grant 465376/2014-2).

This document was prepared by the DES Collaboration using the resources of the Fermi National Accelerator Laboratory (Fermilab), a U.S. Department of Energy, Office of Science, Office of High Energy Physics HEP User Facility. Fermilab is managed by Fermi Forward Discovery Group, LLC, acting under Contract No. 89243024CSC000002.

Part of this research was carried out at the Jet Propulsion Laboratory, California Institute of Technology, under contract with the National Aeronautics and Space Administration (80NM0018D0004).

This research used resources of the National Energy Research Scientific Computing Center (NERSC), a DOE Office of Science User Facility supported by the Office of Science of the U.S. Department of Energy under Contract No. DE-AC02-05CH11231. This work also used resources on the CCAPP condo of the Ruby Cluster at the Ohio Supercomputing Center \cite{OSC1987}, the Duke Compute Cluster (DCC) at Duke University, and the Advanced Research Computing Center at the University of Cincinnati.  Plots in this manuscript were produced partly with \textsc{Matplotlib} \cite{Hunter:2007}, and it has been prepared using NASA's Astrophysics Data System Bibliographic Services.

\section*{Authors' Contribution} \label{sec:contributions}

\begin{itemize}


\item Weak lensing and galaxy clustering analysis (3$\times$2pt) \citepalias{y6-3x2pt}:
M.~Adamow, A.~Alarc{\'o}n, A.~Amon, D.~Anbajagane, W.~d'Assignies, S.~Avila, J.~Beas-Gonzalez, K.~Bechtol, M.~R.~Becker, G.~M.~Bernstein, J.~Blazek, G.~Camacho-Ciurana, G.~Campailla, A.~Campos, A.~Carnero~Rosell, C.~Chang, J.~M.~Coloma-Nadal, M.~Crocce, J.~DeRose, T.~Diehl, A.~Drlica-Wagner, J.~Elvin-Poole, S.~Everett, A.~Fert{\'e}, M.~Gatti, G.~Giannini, R.~A.~Gruendl, W.~G.~Hartley, M.~Jarvis, E.~Krause, S.~Lee, E.~Legnani, S.~Mau, J.~Mena-Fern{\'a}ndez, F.~Menanteau, J.~Muir, J.~Myles, A.~Pieres, A.~Porredon, J.~Prat, M.~Raveri, W.~Riquelme, M.~Rodr{\'i}guez-Monroy, P.~Rogozenski, A.~Roodman, E.~S.~Rykoff, S.~Samuroff, C.~Sanchez, D.~Sanchez-Cid, T.~Schutt, I.~Sevilla-Noarbe, E.~Sheldon, M.~Tabbutt, C.-H.~To, M.~A.~Troxel, J.~de~Vicente, N.~Weaverdyck, A. Whyley, M.~Yamamoto, B.~Yanny, B.~Yin.

\item BAO analysis \cite{y6-baokp}: 
M.~Adamow, S.~Avila, K.~Bechtol, H.~Camacho, R.~Cawthon, K.~C.~Chan, L.~Toribio San Cipriano, M.~Crocce, T.~M.~Davis, T.~Diehl, A.~Drlica-Wagner, J.~Elvin-Poole, I.~Ferrero, G.~Giannini, R.~A.~Gruendl, W.~G.~Hartley, J.~Mena-Fern{\'a}ndez, W.~J.~Percival, A.~Pieres, A.~Porredon, M.~Rodriguez-Monroy, A.~Carnero Rosell, A.~J.~Ross, E.~S.~Rykoff, C.~Sanchez, E.~Sanchez, I.~Sevilla-Noarbe, E.~Sheldon, J.~de Vicente, N.~Weaverdyck, B.~Yanny.

\item Supernova Type Ia analysis \cite{DESY5SN2024}: 
P.~Armstrong, D.~Brout, R.~Camilleri, A.~Carr, R.~Chen, T.~M.~Davis, J.~Frieman, L.~Galbany, K.~Glazebrook, S.~R.~Hinton, R.~Kessler, J.~Lee, C.~Lidman, A.~Möller, B.~Popovic, H.~Qu, M.~Sako, D.~Scolnic, P.~Shah, M.~Smith, M.~Sullivan, B.~O.~Sánchez, G.~Taylor, M.~Toy, M.~Vincenzi, P.~Wiseman.




\item Analysis development and pipeline: O.~Alves, G.~Campailla, A.~Fert{\'e}, K.~Koyama, S.~Lee, A.~Liddle, M.~X.~Lin, J.~Muir, M.~Raveri, J.~Rebouças, P.~Rogozenski, R.~Rosenfeld, D.~Sanchez-Cid, N.~Shah, and D.~Souza. 

\item Parameter estimation, validation, and robustness tests: O.~Alves, G.~Campailla, S.~Lee, J.~Muir, M.~Raveri, and J.~Rebou{\c{c}}as. 

\item Paper writing: O.~Alves, M.~R.~Becker, G.~Campailla, C.~Chang, M.~Crocce, A.~Fert{\'e}, J.~Frieman, K.~Koyama, S.~Lee, A.~Liddle, M.~X.~Lin, J.~Muir, J.~Prat, M.~Raveri, J.~Rebou{\c{c}}as, M.~A.~Troxel. 

\item Coordination and scientific management: 
M.~R.~Becker and M.~Crocce (Year Six Key Project Coordinators), C.~Chang and M.~A.~Troxel (Science Committee Chairs), A.~Alarc{\'o}n, A.~Amon, O.~Alves, S.~Avila, J.~Blazek, A.~Fert{\'e}, G.~Giannini, S.~Lee,  A.~Porredon, J.~Prat, M.~Raveri, M.~Rodr{\'i}guez-Monroy, C.~Sanchez, D.~Sanchez-Cid, T.~Schutt, N.~Weaverdyck, M.~Yamamoto, B.~Yin (Working Group and Analysis Team leads).

\end{itemize}

The remaining authors have made contributions to this paper that include, but are not limited to, the construction of DECam and other aspects of collecting the data; data processing and calibration; developing broadly used methods, codes, and simulations; running the pipelines and validation tests; and promoting the science analysis.

\section*{Data Availability}~\label{sec:data_availability}

The data used in this analysis, including catalogues, datavector measurements, likelihoods and chains, will be made public upon journal acceptance. The DES Y6 Gold catalogue derived from the DES Data Release 2 (DR2) is publicly available at \url{https://des.ncsa.illinois.edu/releases}. 

\bibliographystyle{apsrev4-1}
\bibliography{refs_short, des_short, y1kp, y3kp, y6kp}

\clearpage

\onecolumngrid
\begin{center}
{\large \textbf{Supplemental Material}}
\end{center}

\twocolumngrid

\setcounter{equation}{0}
\setcounter{figure}{0}
\setcounter{table}{0}
\renewcommand{\theequation}{S\arabic{equation}}
\renewcommand{\thefigure}{S\arabic{figure}}
\renewcommand{\thetable}{S\arabic{table}}
\renewcommand{\bibnumfmt}[1]{[S#1]}
\renewcommand{\citenumfont}[1]{S#1}

\begin{table*}
\centering
\renewcommand{\arraystretch}{2.6}
\caption{Summary of marginalized parameter constraints and MAP values (in-between parenthesis) in the $w_0w_a$CDM model.}
\label{tab:full_params}
\resizebox{\textwidth}{!}{%
\begin{tabular}{lcccccccccc}
\hline\hline
Dataset & $S_8$ & $\Omega_{\rm m}$ & $\sigma_8$ & $\Omega_{\rm b}\times10^2$ & $n_s$ & $h$ & $w_0$ & $w_a$ & $w_p$ & $z_p$ \\
\hline
3$\times$2pt & \makecell{$0.810^{+0.023}_{-0.022}$ \\ $(0.834)$} & \makecell{$0.324 \pm 0.031$ \\ $(0.316)$} & \makecell{$0.782^{+0.036}_{-0.040}$ \\ $(0.813)$} & \makecell{$5.1^{+1.1}_{-1.0}$ \\ $(5.6)$} & \makecell{$0.963^{+0.015}_{-0.030}$ \\ $(0.932)$} & \makecell{$0.681^{+0.051}_{-0.081}$ \\ $(0.773)$} & \makecell{$< -1.3$ \\ $95\%~\mathrm{CL}$} & \makecell{$-0.4^{+1.1}_{-0.8}$ \\ $(-0.0)$} & \makecell{$-0.86^{+0.19}_{-0.14}$ \\ $(-0.72)$} & $0.30$ \\
SN + DES BAO & $-$ & \makecell{$0.320^{+0.065}_{-0.041}$ \\ $(0.400)$} & $-$ & \makecell{$4.2^{+0.3}_{-1.2}$ \\ $(3.0)$} & \makecell{$0.965^{+0.019}_{-0.029}$ \\ $(0.948)$} & \makecell{$0.714^{+0.084}_{-0.031}$ \\ $(0.800)$} & \makecell{$-0.80^{+0.12}_{-0.11}$ \\ $(-0.75)$} & \makecell{$-1.1^{+1.2}_{-0.9}$ \\ $(-3.0)$} & \makecell{$-0.87^{+0.10}_{-0.08}$ \\ $(-0.93)$} & $0.06$ \\
\makecell[l]{\textbf{All DES} \\ ($3\times2$pt + SN + DES BAO)} & \makecell{$0.809^{+0.016}_{-0.018}$ \\ $(0.816)$} & \makecell{$0.301^{+0.010}_{-0.012}$ \\ $(0.295)$} & \makecell{$0.808^{+0.027}_{-0.026}$ \\ $(0.822)$} & \makecell{$4.62^{+0.85}_{-0.69}$ \\ $(5.27)$} & \makecell{$0.964^{+0.017}_{-0.029}$ \\ $(0.937)$} & \makecell{$0.708^{+0.071}_{-0.047}$ \\ $(0.791)$} & \makecell{$-0.84 \pm 0.10$ \\ $(-0.82)$} & \makecell{$-0.44^{+0.60}_{-0.55}$ \\ $(-0.44)$} & \makecell{$-0.91 \pm 0.04$ \\ $(-0.90)$} & $0.19$ \\
\makecell[l]{All data \\ (All DES + DESI BAO + CMB)} & \makecell{$0.823 \pm 0.008$ \\ $(0.824)$} & \makecell{$0.311 \pm 0.005$ \\ $(0.312)$} & \makecell{$0.808 \pm 0.008$ \\ $(0.808)$} & \makecell{$4.949^{+0.079}_{-0.083}$ \\ $(4.956)$} & \makecell{$0.974 \pm 0.003$ \\ $(0.974)$} & \makecell{$0.674^{+0.006}_{-0.005}$ \\ $(0.674)$} & \makecell{$-0.82 \pm 0.05$ \\ $(-0.78)$} & \makecell{$-0.63^{+0.21}_{-0.18}$ \\ $(-0.75)$} & \makecell{$-0.97 \pm 0.02$ \\ $(-0.97)$} & $0.32$ \\
\hline
SN + BAO & \makecell{$0.92^{+0.32}_{-0.27}$ \\ $(1.18)$} & \makecell{$0.312^{+0.017}_{-0.012}$ \\ $(0.314)$} & \makecell{$0.90^{+0.31}_{-0.24}$ \\ $(1.16)$} & \makecell{$5.0^{+1.0}_{-0.7}$ \\ $(5.3)$} & \makecell{$0.965^{+0.019}_{-0.028}$ \\ $(0.943)$} & \makecell{$0.689^{+0.060}_{-0.082}$ \\ $(0.713)$} & \makecell{$-0.84^{+0.06}_{-0.08}$ \\ $(-0.84)$} & \makecell{$-0.54^{+0.44}_{-0.48}$ \\ $(-0.55)$} & \makecell{$-0.91^{+0.03}_{-0.04}$ \\ $(-0.91)$} & $0.15$ \\
3$\times$2pt + DESI BAO & \makecell{$0.808^{+0.016}_{-0.015}$ \\ $(0.809)$} & \makecell{$0.337^{+0.022}_{-0.015}$ \\ $(0.356)$} & \makecell{$0.763^{+0.022}_{-0.026}$ \\ $(0.743)$} & \makecell{$5.27^{+0.85}_{-0.80}$ \\ $(5.03)$} & \makecell{$0.961^{+0.013}_{-0.029}$ \\ $(0.938)$} & \makecell{$0.662^{+0.031}_{-0.076}$ \\ $(0.626)$} & \makecell{$-0.61^{+0.24}_{-0.11}$ \\ $(-0.44)$} & \makecell{$-1.17^{+0.47}_{-0.72}$ \\ $(-1.78)$} & \makecell{$-0.95^{+0.06}_{-0.05}$ \\ $(-0.96)$} & $0.41$ \\
CMB + DESI BAO & \makecell{$0.846^{+0.012}_{-0.011}$ \\ $(0.851)$} & \makecell{$0.341^{+0.018}_{-0.011}$ \\ $(0.352)$} & \makecell{$0.794^{+0.012}_{-0.014}$ \\ $(0.786)$} & \makecell{$5.37^{+0.27}_{-0.18}$ \\ $(5.53)$} & \makecell{$0.971 \pm 0.003$ \\ $(0.971)$} & \makecell{$0.647^{+0.009}_{-0.017}$ \\ $(0.637)$} & \makecell{$-0.54^{+0.19}_{-0.07}$ \\ $(-0.43)$} & \makecell{$-1.40^{+0.29}_{-0.50}$ \\ $(-1.67)$} & \makecell{$-1.00 \pm 0.04$ \\ $(-0.99)$} & $0.50$ \\
\hline
\makecell[l]{All DES + DESI BAO \\ (All except CMB)} & \makecell{$0.801^{+0.015}_{-0.014}$ \\ $(0.796)$} & \makecell{$0.313 \pm 0.008$ \\ $(0.310)$} & \makecell{$0.785^{+0.020}_{-0.019}$ \\ $(0.782)$} & \makecell{$4.9^{+1.0}_{-0.6}$ \\ $(5.0)$} & \makecell{$0.965^{+0.021}_{-0.024}$ \\ $(0.967)$} & \makecell{$0.687^{+0.059}_{-0.067}$ \\ $(0.686)$} & \makecell{$-0.84^{+0.06}_{-0.07}$ \\ $(-0.87)$} & \makecell{$-0.53^{+0.33}_{-0.28}$ \\ $(-0.43)$} & \makecell{$-0.94 \pm 0.03$ \\ $(-0.95)$} & $0.22$ \\
\makecell[l]{SN + BAO + CMB \\ (All except $3\times2$pt)} & \makecell{$0.833^{+0.010}_{-0.009}$ \\ $(0.832)$} & \makecell{$0.313^{+0.005}_{-0.006}$ \\ $(0.311)$} & \makecell{$0.816^{+0.010}_{-0.009}$ \\ $(0.817)$} & \makecell{$4.937^{+0.072}_{-0.083}$ \\ $(4.919)$} & \makecell{$0.972 \pm 0.003$ \\ $(0.972)$} & \makecell{$0.675^{+0.006}_{-0.005}$ \\ $(0.676)$} & \makecell{$-0.80^{+0.05}_{-0.06}$ \\ $(-0.81)$} & \makecell{$-0.72^{+0.22}_{-0.20}$ \\ $(-0.72)$} & \makecell{$-0.97 \pm 0.02$ \\ $(-0.97)$} & $0.30$ \\
\makecell[l]{3$\times$2pt + BAO + CMB \\ (All except SN)} & \makecell{$0.834^{+0.010}_{-0.009}$ \\ $(0.829)$} & \makecell{$0.337^{+0.018}_{-0.012}$ \\ $(0.330)$} & \makecell{$0.788^{+0.012}_{-0.015}$ \\ $(0.791)$} & \makecell{$5.34^{+0.28}_{-0.20}$ \\ $(5.25)$} & \makecell{$0.973 \pm 0.003$ \\ $(0.974)$} & \makecell{$0.650^{+0.010}_{-0.017}$ \\ $(0.655)$} & \makecell{$-0.57^{+0.20}_{-0.10}$ \\ $(-0.62)$} & \makecell{$-1.25^{+0.33}_{-0.51}$ \\ $(-1.10)$} & \makecell{$-1.00 \pm 0.04$ \\ $(-1.00)$} & $0.53$ \\
\makecell[l]{3$\times$2pt + SN + CMB \\ (All except BAO)} & \makecell{$0.823 \pm 0.008$ \\ $(0.820)$} & \makecell{$0.305 \pm 0.007$ \\ $(0.302)$} & \makecell{$0.816^{+0.011}_{-0.010}$ \\ $(0.817)$} & \makecell{$4.84^{+0.10}_{-0.12}$ \\ $(4.81)$} & \makecell{$0.973 \pm 0.003$ \\ $(0.974)$} & \makecell{$0.682 \pm 0.008$ \\ $(0.684)$} & \makecell{$-0.76^{+0.08}_{-0.09}$ \\ $(-0.80)$} & \makecell{$-0.95^{+0.42}_{-0.38}$ \\ $(-0.81)$} & \makecell{$-0.96^{+0.02}_{-0.03}$ \\ $(-0.96)$} & $0.25$ \\
\makecell[l]{All DES + CMB \\ (All except DESI BAO)} & \makecell{$0.822^{+0.007}_{-0.008}$ \\ $(0.821)$} & \makecell{$0.303 \pm 0.007$ \\ $(0.301)$} & \makecell{$0.819^{+0.010}_{-0.009}$ \\ $(0.820)$} & \makecell{$4.81^{+0.10}_{-0.11}$ \\ $(4.79)$} & \makecell{$0.973 \pm 0.003$ \\ $(0.974)$} & \makecell{$0.684^{+0.007}_{-0.008}$ \\ $(0.686)$} & \makecell{$-0.74 \pm 0.08$ \\ $(-0.72)$} & \makecell{$-1.10^{+0.42}_{-0.36}$ \\ $(-1.17)$} & \makecell{$-0.96^{+0.03}_{-0.02}$ \\ $(-0.95)$} & $0.25$ \\
\hline\hline
\end{tabular}
}
\end{table*}

\subsection{Implications for Growth of Structure}

\begin{figure}
    \centering
    \includegraphics[width=\columnwidth]{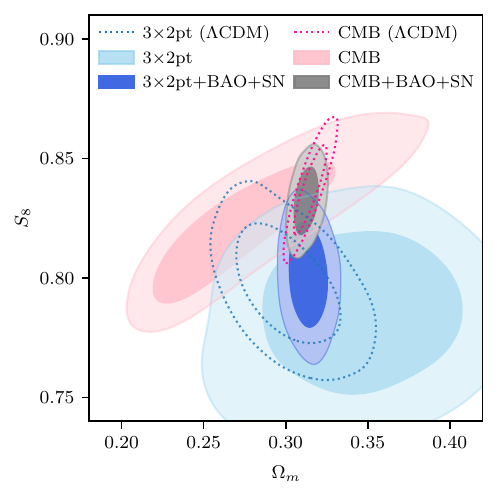}
    \caption{
    Constraints on $\Omega_m$ and $S_8$ under $\Lambda$CDM and $w_0w_a$CDM. Dotted contours show $\Lambda$CDM constraints from the CMB (pink) and 3$\times$2pt (light blue). Filled contours show the corresponding $w_0$--$w_a$ results: CMB alone (pink filled) and 3$\times$2pt alone (light blue filled) exhibit large degeneracies, and the formal tension is reduced. Adding BAO + SN to anchor the expansion history yields CMB + BAO + SN (gray) and 3$\times$2pt + BAO + SN (dark blue). 
    } 
    \label{fig:s8tension}
\end{figure}

The DES Y6 3$\times$2pt $\Lambda$CDM results provide some of the tightest constraints on growth of structure in the low-redshift universe \citepalias{y6-3x2pt}. 
Comparing the inferred $S_8$ to high-redshift predictions from the primary CMB is a powerful test of the concordance cosmology model. In  Ref.~\citepalias{y6-3x2pt}, we found that 3$\times$2pt and the primary CMB have a parameter difference at the 2.2$\sigma$ level in the $\Omega_{\rm m}$--$S_8$ plane, and $2.6\sigma$ in the $S_8$ direction. With our baseline analysis choices and assuming a 
$\Lambda$CDM model, we obtain $1.9\sigma$ and $2.0\sigma$, respectively. 
The differences relative to values reported in Ref.~\citepalias{y6-3x2pt} reflect our different analysis choices described above, and are within the range validated by our robustness tests \cite{y6-extensions}.

There have been many similar tests in the literature that have found 1--3$\sigma$ differences~\cite{y1-keypaper, y3-3x2ptkp, y3-harmonic, DESKIDS2023, kids-1000-shear, kids-1000-improved, kids-1000-boss, kids-legacy, hsc-y3-shear, hsc-y3-shear-harmonic, hsc-y3-3x2, hsc-y3-3x2-emu,y3-3x2pt-cluster} -- this is often discussed under the term `$S_8$ tension.' Here, we are interested in whether the overall picture changes when we move to a dynamical dark energy model.

Figure~\ref{fig:s8tension} shows the $\Omega_{\rm m}$--$S_8$ constraint from 3$\times$2pt and the primary CMB under $\Lambda$CDM and $w_0w_a$CDM.
Since the $w_0w_a$CDM model has more degrees of freedom in the late-time expansion history, the constraining power on $\Omega_{\rm m}$ and $S_8$ with only CMB or 3$\times$2pt is reduced. 
The larger error bars reduce the offset to 0.4$\sigma$ in $S_8$ and 1.1$\sigma$ in the $\Omega_{\rm m}$--$S_8$ plane.
The constraints on the growth parameters are
\twolineconstraint
{\Omega_\textrm{m} &= 0.32 \pm 0.03}
{S_8 &= 0.81 \pm 0.02}
{3$\times$2pt  (68\% C.L.),}
\noindent consistent with the $\Lambda$CDM result as shown in Fig.~\ref{fig:s8tension}.

We further examine this difference when additional geometric anchors from BAO and SN are included.
The geometric data provide tighter constraints on $\Omega_{\rm m}$ and break the degeneracy between $S_8$ and $\Omega_{\rm m}$. 
Under this common anchoring, we compare constraints from CMB + BAO + SN and 3$\times$2pt + BAO + SN. 
The two combinations show parameter difference at 1.2$\sigma$ in the $\Omega_{\rm m}$--$S_8$ plane and 1.7$\sigma$ in $S_8$ (see Table \ref{tab:full_params} in the Supplemental Material for full parameter constraints). We note that in $\Lambda$CDM, the same geometric anchoring results in a $0.7\sigma$ difference in the $\Omega_{\rm m}$--$S_8$ plane and $1.1\sigma$ in $S_8$.
This suggests that the overall consistency between the growth parameters in the early and late Universe remain unchanged under $w_0w_a$CDM.

\subsection{Marginalized parameter constraints}
In Table \ref{tab:full_params} we summarize the marginalized parameter constraints and MAP values obtained from the various dataset combinations explored in this paper in the $w_0 w_a$CDM model. 

\end{document}